\documentclass[letter]{aa}
\usepackage{natbib}
\usepackage{latexsym}
\usepackage{amsmath}
\usepackage{amssymb}
\usepackage{graphicx}
\usepackage[english]{babel}
\newcommand{\el}[2]{\ensuremath{\rm^{#2}\kern-0.8pt\rm#1}}
\begin{document}

\title{A new imprint of fast rotators: low \boldmath \el{C}{12}/\el{C}{13}
  ratios in extremely metal-poor halo stars}

\author{C. Chiappini\inst{1,2}, S. Ekstr\"om\inst{1},
  G. Meynet\inst{1}, R. Hirschi\inst{3}, A. Maeder\inst{1}, \and C. Charbonnel\inst{1,4}}

\institute{Observatoire Astronomique de l'Universit\'e de Gen\`eve, CH-1290,
Sauverny, Switzerland
\and Osservatorio Astronomico di Trieste, Via G. B. Tiepolo 11, I - 34131 Trieste, Italia
\and Keele University, Lennard-Jones Lab., Keele, ST5 5BG, UK
\and LATT, CNRS UMR 5572,
OMP, Universit\'e Paul Sabatier Toulouse 3, 14 Av. E. Belin, 31400
Toulouse, France}

\date{Received / Accepted}

\abstract
{Fast stellar rotation is currently the most promising mechanism for producing primary nitrogen in metal-poor massive stars. Chemical evolution models computed with the inclusion of the yields of fast rotating models at a metallicity $Z=10^{-8}$ can account for the high N/O abundances observed in normal metal-poor halo stars. If, as believed, intermediate mass stars did not have enough time to contribute to the interstellar medium enrichment at such low metallicities, the above result constitutes a strong case for the existence of fast rotators in the primordial Universe.}
{An important result of stellar models of fast rotators is that large quantities of primary \el{C}{13} are produced. Hence, our goal is to investigate the consequence of fast rotation on the evolution of the \el{C}{12}/\el{C}{13} ratio in the interstellar medium at low metallicity.}
{We compute the evolution of the \el{C}{12}/\el{C}{13} ratio for the first time at very low metallicities upon the inclusion of fast rotators at $Z= 10^{-8}$.}
{We predict that, if fast rotating massive stars were common phenomena in the early Universe, the primordial interstellar medium of galaxies with a star formation history similar to the one inferred for our galactic halo should have \el{C}{12}/\el{C}{13} ratios between 30-300. Without fast rotators, the predicted \el{C}{12}/\el{C}{13} ratios would be $\sim 4500$ at [Fe/H] $= -3.5$, increasing to $\sim 31000$ at around [Fe/H] $= -5.0$. Current data on very metal-poor giant normal stars in the galactic halo agree better with chemical evolution models including fast rotators. The expected difference in the \el{C}{12}/\el{C}{13} ratios, after accounting for the effects of the first dredge-up, between our predictions with/without fast rotators is of the order of a factor of 2-3.  However, larger differences (a factor of $\sim 60-90$) are expected for giants at [Fe/H]$=-5$ or turnoff stars already at [Fe/H]$=-3.5$. To test our predictions, challenging measurements of the \el{C}{12}/\el{C}{13} in more extremely metal-poor giants and turnoff stars are required.}
{}


   \authorrunning{C. Chiappini et al.}

   \titlerunning{A new imprint of fast rotating stars at low metallicities}

\maketitle

\section{Introduction}

In the last few years, a burst of new information on cosmic abundances, provided by high-resolution spectra of very metal-poor stars in the halo has had a considerable impact on our understanding of both stellar nucleosynthesis and the chemical enrichment of the primordial interstellar medium (ISM). In particular, it is in the halo of the Milky Way that the oldest and most metal-poor stars in the Universe are observable, born at times (or equivalent redshifts) still out of reach of the deepest surveys of primordial galaxies. 

Several teams \citep[e.g.][]{cayrel04,cohen07,bark05,aok07} have studied the chemical composition of halo metal-poor stars (stars with less than 1/300 down to 1/10000 the solar iron abundance). These low-mass stars have lifetimes comparable to the age of the Universe, and they retain in their atmospheres the elemental abundances of the gas at the time of their birth. Hence, these stars contain a memory of the unique nucleosynthesis contribution of the first stellar generations to the enrichment of the ISM, offering a local benchmark for cosmology. These massive stars are long dead but there are hopes of observing them directly in very high-redshift galaxies. The current way to constrain the properties of the first generations of stars and check whether the very different primordial environment has produced noticeable effects concerning their properties is to search for their imprints on the oldest extremely metal-poor stars (EMPs) in our galactic halo.

Recently a sample of EMP ([Fe/H]$\leq -3$) halo stars has become available. Using UVES on the VLT, \citet{cayrel04} provided abundance measurements of unprecedented accuracy for several elements in a fairly large sample of metal-poor \emph{normal} stars \citep[to distinguish from the C-rich EMP stars, see][]{beers05}. Their data have revealed a striking homogeneity in the chemical properties of halo stars. In particular, a very low scatter for the [$\alpha$/Fe] ratios was found. In a subsequent paper, the authors \citep[, hereafter S05]{spite05} again report some unexpected results, namely: these same metal-poor stars show a high N/O ratio suggesting high levels of production of primary nitrogen in massive stars (see below). Furthermore, a large scatter in their N/O ratios (much larger than their quoted error bars) was found. S05 also report a slight increase in the C/O of the same stars with decreasing metallicity.

As discussed in \citet{chiap05,chiap06a,chiap06b}, the data of S05 happen to be in a very interesting metallicity range: at such low metallicities, their observed stars are probably made of only massive star ejecta diluted by the primordial ISM. Indeed, chemical evolution models (CEM) of the halo \citep{chiap06a,pran03} show that an 8 $M_{\sun}$ star dies at [Fe/H] $\sim -3.5$. This means that, below this metallicity, the ISM is enriched exclusively by massive stars. Moreover, according to these models the contribution of AGB stars is negligible below [Fe/H] $\sim -2.5$. More important, these theoretical predictions seem to be confirmed by very recent observations. \citet{mel07} have shown that the $^{25,26}$Mg/$^{24}$Mg ratios in halo dwarfs are low and that AGB stars would have played a minor role below [Fe/H] $\sim -2.0$.

If AGB stars\footnote{Classically the best candidates for the production of primary N through hot-bottom burning during the asymptotic giant branch (AGB) phase \citep{siess07}.} indeed had not had enough time to contribute to the ISM enrichment at such low metallicities and massive stars are not producers of primary nitrogen \citep[as predicted by standard stellar models, e.g.,][]{whw02}, one would expect to observe a decline in the N/O or N/Fe ratios towards low $Z$, contrary to what has been found by S05. 
Hence, the high levels of N/O observed in halo stars have to be a result of the nucleosynthesis taking place in the metal-poor massive stars, suggesting a revision of standard models for the nucleosynthesis
in massive stars.

The effects of stellar axial rotation are numerous and at low metallicity may lead to a drastic revision of current wisdom \citep[see][]{mem06,hir07}. The stellar models of the Geneva group, including rotation \citep{mm00,mm01}, have proved to be successful in explaining some observations that could not be explained by non-rotating models, namely, the observed number ratio of Wolf Rayet stars to O-type stars for different metallicities, the observed ratio of WN to WC for metallicities lower than solar, and the observed ratio of type Ib/Ic to type II supernovae at different metallicities. One of the consequences of rotation is that carbon and oxygen, produced in the He-burning core, are transported by rotational mixing into the H-burning shell, where they are transformed into primary \el{C}{13} and \el{N}{14}. Interestingly, the efficiency of this process increases when the initial mass and rotational velocity increase \citep[see][]{mm00,mm02b}, producing more N for fast rotators. Another fundamental prediction from models including rotation is that the rotational mixing also increases with decreasing metallicity.

\citet{chiap06b} computed CEMs adopting the new calculations of \citet{hir07} for the evolution of massive stars at very low metallicities under the assumption of an almost constant initial ratio $\upsilon_{\rm rot}/\upsilon_{\rm crit}$ as a function of metallicity\footnote{$\upsilon_{\rm crit}$ represents the critical rotational velocity, i.e., the equatorial velocity such that the centrifugal acceleration compensates for the gravity.} (i.e. where the $\upsilon_{\rm rot}$ increases towards lower metallicities). It was shown that, in such a framework, massive stars can produce large amounts of nitrogen, and this explains the data of S05. Moreover, these same models naturally predict a C/O upturn at low metallicities. It should be noticed that this is currently the only way to explain the CNO halo abundances of \emph{normal} stars giving strong support to the idea that stars rotate faster at low metallicities. One important caveat is that the abundances of C, N, and O in low metallicity red giant branch stars, could still suffer from systematic non negligible corrections due to 3D and NLTE effects \citep[see][]{aspl05}. Are there other imprints of fast rotators that are less affected by the above uncertainties? The \el{C}{12}/\el{C}{13} ratio is largely unaffected by uncertainties in the adopted atmosphere parameters \citep[, hereafter S06]{spite06}, although in this case the stellar evolution effects taking place in these old, low-mass stars have to be correctly accounted for in order to recover their pristine values.

In this Letter we suggest that fast-rotating massive stars could also have left an imprint on the evolution of the \el{C}{12}/\el{C}{13} ratio in the earliest phases of the ISM enrichment. In fact, the same rotational mixing responsible for the nitrogen production would also produce non negligible amounts of \el{C}{13}.
We predict, for the first time, that the impact of an early population of fast rotators on the evolution of the
\el{C}{12}/\el{C}{13} abundance ratio in the very metal-poor ISM is huge. \emph{If fast rotators existed, low ($\sim 30-300$) \el{C}{12}/\el{C}{13} ratios should be observed in the very metal-poor ISM of galaxies polluted essentially only by massive stars.}

This Letter is organized as follows. The adopted stellar yields are briefly presented in Sect.~\ref{syields}. Our results are shown in Sect.~\ref{sres}. Section~\ref{sdiscu} is devoted to our conclusion and to a discussion of this finding's impact on different research areas.

\section{\boldmath $^{13}$C yields for very metal-poor stars \label{syields}}

As this Letter focuses on the very metal-poor range (below [Fe/H] $\simeq -3$ or log(O/H)+12 $\simeq 7$) where no/few intermediate mass star would have had time to contribute to the ISM enrichment \citep[see][]{mel07}\footnote{The effect of rotation on the evolution of low- and intermediate-mass stars (LIMS) and their impact on the galactic ISM will be discussed in a forthcoming paper. In the present work the low- and intermediate-mass stars are included with the same yield prescriptions as adopted in our previous work \citep{chiap03}, with $\upsilon^{\rm ini}_{\rm rot}=300$\,km\,s$^{-1}$.}, we concentrate on the effects of rotation on the stellar yields of metal-poor massive stars.

The $^{13}$C is produced in H-burning regions, where the CN cycle converts $^{12}$C into $^{13}$C. As discussed in the previous section, rotation triggers the production of primary \el{C}{13}, by allowing the diffusion of \el{C}{12} produced in He-burning zones into H-burning ones. As a numerical illustration, a 40 $M_{\sun}$ at $Z=10^{-8}$ ejects a quantity of newly synthesized \el{C}{13} equal to $6.3 \times 10^{-2}\ M_{\sun}$ when $\upsilon^{\rm ini}_{\rm rot} =$700 km\,s$^{-1}$ \citep{hir07}. For comparison, models without rotation predict a \el{C}{13} yield of $3.3 \times 10^{-5}\ M_{\sun}$ for $Z= 10^{-6}$ and $M=35\ M_{\sun}$ \citep{chief04}, or a \el{C}{13} yield of $2.2 \times 10^{-4}\ M_{\sun}$ for $Z=0.001$ and $M=40\ M_{\sun}$ \citep{tom07}.

Here we adopt the \el{C}{12} and \el{C}{13} yields of \citet{hir07} computed for $Z=10^{-8}$ and initial rotational velocities, $\upsilon^{\rm ini}_{\rm rot}$=600 to 800\,km\,s$^{-1}$, depending on the stellar mass. For higher metallicities the adopted yields are from \citet[, hereafter MM02]{mm02a}, for $\upsilon^{\rm ini}_{\rm rot}$=300\,km\,s$^{-1}$. The justification for the higher velocities at $Z=10^{-8}$ relies on the following assumption: stars begin their evolution on the ZAMS with a rotational velocity such that the ratio $\upsilon_{\rm rot}/\upsilon_{\rm crit}$ remains almost constant (around 0.5-0.6) with mass and metallicity, implying higher rotational velocities at lower metallicities. As a consequence the stellar yields of \el{C}{13} of \citet{hir07} are greater by 2 to 3 orders of magnitude than the ones of MM02 for $Z= 10^{-5}$ and $\upsilon^{\rm ini}_{\rm rot}$=300\,km\,s$^{-1}$.

\section{Results \label{sres}}

Here we show the impact of the stellar yields discussed above on the predictions of the \el{C}{12}/\el{C}{13} evolution in the galactic halo. We adopt the same CEM of \citet{chiap06a}, briefly described below.

Our model assumes that the galactic halo was formed by a combination of infall plus outflow. Our main assumptions are: a) a Gaussian infall ($f(t) \propto e^{(t-t_0)^2/2 \sigma^2}$, with $t_0 = $0.1 Gyr and $\sigma =$ 0.05); b) an outflow rate of 8 times the star formation rate; c) a Schmidt law for the star formation rate \citep[see also][]{pran03}. This model is able to reproduce the observed halo metallicity distribution \citep[see Fig.~4 of][]{chiap06a}. While in our previous model, without outflow \citep{chiap06b} a 8 $M_{\sun}$ star would die at a metallicity [Fe/H]$\sim -2.2$, in the new model with outflow the same star would die when the metallicity in the ISM was [Fe/H]$\sim -3.8$. The latter model is favoured by the observed halo metallicity distribution.

\begin{figure}
\centering
\includegraphics[width=5.5cm,angle=0]{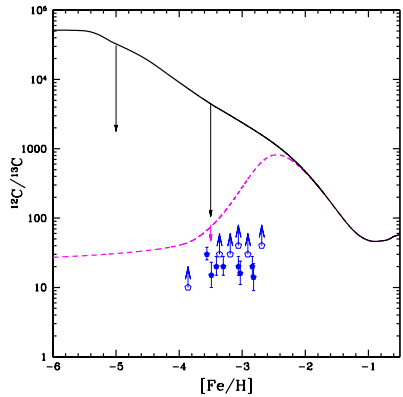}
\caption{Predicted evolution of the \el{C}{12}/\el{C}{13} ratio according to a CEM computed with different stellar yield sets for metallicities below $Z=10^{-5}$: a) solid line - a model computed under the assumption that the lowest metallicity yield table of MM02 ($Z=10^{-5}$) is valid down to $Z=0$ (without fast rotators) and b) dashed line (magenta in the on-line version) a model computed with the contribution of $Z=10^{-8}$ fast rotators. The data are the \emph{unmixed} stars of S06. The open symbols represent lower limits. The arrows indicate the final \el{C}{12}/\el{C}{13} observed in giants, starting from the initial values given by models with (light/magenta arrow) and without fast rotators (dark arrows).}
\label{fig1}
\end{figure}

In Fig.~\ref{fig1} we show our model predictions. The different curves show the same CEM computed with different sets of stellar yields at low metallicities. A model computed without fast rotators predicts very high $^{12}$C/$^{13}$C ratios at low metallicities, namely: $\sim 2400$ at [Fe/H] $= -3.0$, $\sim 4500$ at [Fe/H] $= -3.5$, increasing to $\sim 31000$ around [Fe/H] $= -5.0$. When the contribution of the fast rotators is taken into account, much lower $^{12}$C/$^{13}$C ratios are predicted: $\sim 295$ at [Fe/H] $= -3.0$, $\sim 80$ at [Fe/H] $= -3.5$ and $\sim 30$ at [Fe/H] $= -5.0$. In other words, we predict smaller ISM $^{12}$C/$^{13}$C ratios (by a factor of $\sim 1000$ at [Fe/H] $= -5.0$ and of $\sim 60$ at [Fe/H] $= -3.5$) upon the inclusion of fast rotators. We notice that the impact of fast rotators could be even greater than what has been computed here. This would be the case if, for instance, fast rotators would still play some role at slightly higher metallicities. However, at present, stellar yields computed with high rotational velocities for higher metallicities ($Z\sim 10^{-5}$) are not available.

Currently, the only data available in the literature for the $^{12}$C/$^{13}$C ratio at such low metallicities is from S06. They obtained new $^{12}$C/$^{13}$C isotopic ratios for the same stars as studied in S05. Here we only plot their so-called \emph{unmixed} stars, which are giants that have undergone the first dredge-up but are still below the luminosity function bump and hence have not experienced the thermohaline-mixing \citep[e.g.,][]{charb07}. To recover their initial $^{12}$C/$^{13}$C ratio relevant to the comparison with our CEM predictions, we have computed the evolution of the carbon isotopic ratio in stellar models of 0.85~$M_{\sun}$, [Fe/H]~$=$~$-3.5$ (typical of the Spite sample) from the pre-main sequence up to the end of the first dredge-up on the red giant branch, assuming different initial values for the $^{12}$C/$^{13}$C ratio.

Starting from an initial \el{C}{12}/\el{C}{13} ratio corresponding to the prediction of a CEM without fast rotators at [Fe/H]$=-3.5$, we obtain a post dredge-up value of 106. This value is clearly higher than the observed ratio in the Spite \emph{unmixed sample} ($\sim 20-30$). In the case of an initial value of 80 as predicted by a model that takes the contribution of the fast rotators at this same metallicity into account, a value of 47 is obtained, i.e., slightly higher than, but marginally consistent with, the observational data. Starting from an initial value of 50 (see previous discussion), the observational values can be reached ($\sim 30$). Thus, even though our predictions for the ISM \el{C}{12}/\el{C}{13} ratio with and without fast rotators differ by a factor of 60 at [Fe/H]$=-3.5$, in the case of very metal-poor giants, the expected effect would be smaller (a factor of 2-3), but still significant. Larger differences are obtained at lower metallicities. At [Fe/H]$=-5$, starting from an initial value of 31000 (predicted by CEM without fast rotators), we obtain a post dredge-up value of 1790 (Fig.~\ref{fig1}). On the other hand, essentially no change is obtained when starting from the initial values of our CEM which include fast rotators (in the latter case the initial \el{C}{13} is already so high that the effect of the 1st dredge-up turns out to be negligible). Thus, at such low metallicities the expected effect is greater (a factor of 60!). 

In summary, we offer a robust theoretical prediction both for the ISM of primordial galaxies and for the EMPs in our galactic halo. Our predictions could in principle be checked, once C-isotopic ratios had been measured in normal EMP turnoff stars at [Fe/H] $\sim -3.5$ or giants at lower metallicities.  An additional test of our predictions will probably come from the fast evolving field of abundance measurements in the ISM of high-redshift galaxies \citep{pett06}. The effect of fast rotators should appear in systems where the main contributors to the gas enrichment are the massive stars. From an observational point of view, the most suitable systems for measuring the \el{C}{12}/\el{C}{13} are the damped Ly$\alpha$ systems (DLAs) because the absence of strong galactic winds would not allow the measurement of the isotopic shifts. \citet{lev06} reported a lower limit \el{C}{12}/\el{C}{13}~$>$~80 in a DLA at z$=1.15$. However in this case, the metallicity of the system is [Fe/H]$\sim -1$ and the role of fast rotators could be minor. The value found by Levshakov et al. cannot be compared to our models in Fig.~\ref{fig1} because the chemical evolution of DLAs most probably proceeded in a radically different way than for the galactic halo \citep[see][]{chiap03,dess07}. The impact of fast rotators in DLAs will be the subject of a forthcoming paper.

It is worth noticing that \el{C}{12}/\el{C}{13} ratios have recently been measured \citep{carr05} in globular cluster unevolved stars, in the metallicity range [Fe/H]$\sim -2.0$ to  $\sim -0.8$. A low value for this ratio was found ($\sim 8$), suggesting the pollution of the studied stars by previously evolved AGB stars. More recently, \citet{decr07} have suggested that the low \el{C}{12}/\el{C}{13} values (together with many other observed properties of globular clusters) could be explained better by instead assuming that the stars we observe today in globular clusters were polluted by a previously evolved generation of fast-rotating massive stars. 

A final remark concerns the contribution of super-AGB stars (SAGB), which in principle could enrich the ISM in \el{C}{13} and \el{N}{14} \citep{siess07,gilp07}, but were not included in the present calculations as the stellar yields for these objects are currently not available. As SAGBs have masses between 8-12 $M_{\sun}$ (depending on the metallicity) one could wonder if these objects could enrich the ISM early on both in \el{C}{13} and \el{N}{14}, thus being good candidates to explain the very metal-poor data on CNO, without the need of fast rotators. However, this does not seem to be the case as the very metal-poor data of Spite et al. falls in a metallicity range were the main contributors to the ISM enrichment are stars with masses between 10 and 30 $M_{\sun}$ provided our enrichment timescale, constrained by the halo metallicity distribution, is correct \citep[see][]{chiap06a}. We should however keep this caveat in mind and this point certainly deserves further study once stellar yields for SAGBs become available.

\section{Discussion and conclusions \label{sdiscu}}

In this Letter we have studied the impact of fast rotators on the chemical evolution of the \el{C}{12}/\el{C}{13} ratio in the halo of the Milky Way. We predict that, if fast rotating massive stars were common phenomena in the early Universe, the ISM at an epoch where the chemical enrichment was essentially due to massive stars ([Fe/H]$<-3.0$) should have had \el{C}{12}/\el{C}{13} ratios between 30-300. Without fast rotators, the predicted \el{C}{12}/\el{C}{13} ratios would be $\sim 4500$ at [Fe/H] $= -3.5$, increasing to $\sim 31000$ around [Fe/H] $= -5.0$. 

We also computed the expected surface abundance modifications during the evolution of a typical, very metal-poor, low mass star, and concluded that if fast rotators did play a role (i.e. the ISM in the earliest phases had low \el{C}{12}/\el{C}{13} ratios), the EMPs observed today ([Fe/H]$\lesssim -3.0$) should 
have similar \el{C}{12}/\el{C}{13} ratios to the ones reported by S06, whereas values a factor of $\sim 2-3$ higher would be obtained without fast rotators. A stronger effect is predicted at lower metallicities where the two different scenarios would lead to differences of about 60.

If the first stars were confirmed to be fast rotators, achieving initial rotational velocities of about 600-800\,km\,s$^{-1}$, the impact on several other research areas will be enormous. The high CNO
content on the surface of very metal-poor fast rotators could be a key to produce the cooling needed to form low-mass stars, even at the very first phases of the Universe. A profound impact is to be expected on the progenitors of long-duration gamma-ray bursts \citep{yoon06,hir07}. Stars born with such high velocities may have their supernovae explosion affected by the fast rotation of the core. 

Our results on the CNO nucleosynthesis are also crucial in interpreting the emission line spectra of star-forming galaxies at high redshifts. For instance \citet{pett02} measured abundances of several elements in the ambient interstellar medium of the Lyman break galaxy MS1512-cB58 and showed that, while O and other $\alpha$-elements have already reached $\sim 2/5$ of their solar values, the production and mixing within the ISM of Fe-peak and N apparently lagged behind, indicating an age of 
a few tenths of Myrs \citep{matt02}. Although this galaxy has a much higher metallicity ($\sim -0.8$) than the ones considered in the present work, it is interesting to notice that if the timescale for the release of primary N changes (as it would be the case if fast rotators play a significant role), the conclusions drawn from high-z emission line observations can be affected.

In addition to nitrogen and carbon, fast rotation also affects other elements, such as \el{Ne}{22}, \el{Mg}{25}, \el{Mg}{26}, or the weak s-process elements, to cite only a few of them. The discussion of these other elements will appear in a future paper.

\begin{acknowledgements}
C. Chiappini acknowledges partial financial support from PRIN INAF 2006 cram 1.06.09.10 and from \emph{Fonds National Suisse de la Recherche Scientifique}. We thank L. Siess, F. Matteucci, and M. Pettini for interesting discussions and the referee whose criticism contributed to improving this work.
\end{acknowledgements}

\bibliographystyle{aa}
\bibliography{8698bib}

\end{document}